\documentclass[A4,twoside,12pt,reqno]{article}

\usepackage[super,compress]{cite}
\usepackage{graphicx}
\usepackage{float}
\usepackage[latin1]{inputenc}

\usepackage{amssymb}
\usepackage{amsthm}
\usepackage[tbtags]{amsmath}
\usepackage{doc}
\usepackage{latexsym}
\usepackage{amscd}
\usepackage[frame,cmtip,arrow,matrix,line,graph,curve]{xy}
\usepackage{graphics}
\usepackage{epsfig}
\usepackage{eucal}
\usepackage{mathrsfs}
\usepackage[margin=1.3in]{geometry}

\begin{document}
\thispagestyle{plain}
 \markboth{}{}
\small{\addtocounter{page}{0} \pagestyle{plain}

\noindent\centerline{\Large \bf On Bianchi type III Cosmological Model  with Quadratic}\\ \centerline{\Large \bf EoS in Lyra Geometry }\\

\noindent\centerline{$^1$Mahabubur Rahman Mollah}\\
\noindent\centerline{$^2$Kangujam Priyokumar Singh}\\
\noindent\centerline{$^3$Pheiroijam Suranjoy Singh}\\
\textbf{}
\noindent\centerline{$^1$Department of Mathematics, Commerce College,}\\
\noindent\centerline{Kokrajhar, Assam - 783370, India }\\
\noindent\centerline{mr.mollah123@gmail.com}\\
\noindent\centerline{$^2$$^,$$^3$Department of Mathematical Sciences, Bodoland University,}\\
\noindent\centerline{Kokrajhar, Assam-783370, India.}\\
\noindent\centerline{$^2$pk\_mathematics@yahoo.co.in}
\noindent\centerline{$^3$surphei@yahoo.com}\\
\\
\footnote{Preprint of an article published in [Int. J. Geom. Methods Mod. Phys., Vol. 15, No. 11, 1850194 (2018)DOI: 10.1142/S0219887818501943]\textsuperscript{\textcopyright}[World Scientific Publishing Company] [Journal URL: www.worldscientific.com/worldscinet/ijgmmp]}
\noindent{\bf Abstract.} The paper deals with the investigation of a homogeneous and anisotropic space-time described by Bianchi type III metric with perfect fluid in Lyra geometry setting. Exact solutions of the Einsten's field equations have been obtained under the assumption of quadratic equation of state (EoS) of the form $p = a \rho^{2} - \rho$ , where $a$ is a constant and strictly $a > 0$. The physical and geometrical aspects are also examined in details.
\noindent{\bf 2010 Mathematics Subject 2010:} 83CXX, 83F05, 83C15.
\vspace{0.2in}

\begin{abstract}
The paper deals with the investigation of a homogeneous and anisotropic space-time described by Bianchi type III metric with perfect fluid in Lyra geometry setting. Exact solutions of the Einsten's field equations have been obtained under the assumption of quadratic equation of state (EoS) of the form $p = a \rho^{2} - \rho$ , where $a$ is a constant and strictly $a > 0$. The physical and geometrical aspects are also examined in details.
\end{abstract}

\section{Introduction}

According to Binney and Tremaine [1], the process of transformation of the universe may be divided into four main eras, which are vacuum energy era (or, Planck era), radiation era $\&$ matter dominated era and the dark energy era (or, de Sitter era). The universe goes through a stage of early inflation in the vacuum energy era (Planck era),  that brings the universe from the Planck size $l_{p} = 1.62 \times 10^{-35} m$ to a 'macroscopic' size $a \sim 10^{-6} m$ in a lowest possible fraction of a second [2, 3]. After that, the universe moves into the radiation era and then, the matter-dominated era, whenever the temperature drops down below 103 K approximately [4]. Finally, in the dark energy era, the universe goes through a phase transition of late inflation [5]. The singularity problems are solved in the early inflation [2] whereas late inflation is necessary to explain accelerating expansion of the universe [6-14]. The cosmological results and data sets like Atacama Cosmology Telescope (ACT) [15], Planck 2015 results- XIII [16], are also suggested about late time inflation of the universe and allows the researchers to determine cosmological parameters such as the Hubble constant $H$ and the deceleration parameter $q$. Studying the constraints given by the data from CMBR (Cosmic Microwave Background Radiation) investigation [17, 18], WMAP (Wilkinson Microwave Anisotropy Probe) [19], observations of clusters of galaxies at low red-shift [20, 21] etc., it can be deduced that the universe is dominated by some mysterious components. This mystical component of the energy is called dark energy which has negative pressure and positive energy density (giving negative EoS parameter). In the energy budget, it is believed that about $73 \%$ of our universe is Dark energy, about $23 \%$ is occupied by Dark matter and the usual baryonic matter occupies about $4 \%$. Therefore, the study of the aspect of dark energy has become an interesting topics in the field of fundamental physics [22-28]. Einstein's cosmological constant $\Lambda$ is the best match for dark energy and physically, it corresponds to the quantum vacuum energy. The cosmological model with $\Lambda$ and cold dark matter (CDM) is usually called the $\Lambda$CDM model.\\
\\
In 1919, Einstein geometrizes gravitation in his theory of general relativity which is treated as a basis for a model of the universe. After that, many cosmologists and astrophysicists attempted to study gravitation in different contexts. For the illustration both gravitation and electromagnetism, Weyl [29] tried to formulate a novel gauge theory containing metric tensor. But, due to the non-integrability condition, it does not get importance in the cosmological society. In order to eliminate the non-integrability condition in the Weyl's geometry, Lyra [30] introduced a gauge function into the structure less manifold and recommended an adjustment in it. These customized theories of relativity are termed as the alternate theories of the gravitation or modified theories of the gravitation. Some significant customized theories of gravitation are Brans-Dicke theory [31], Scalar-tensor theories [32], Nordvedt [33], Saez and Ballester [34], Vector-tensor theory [35], Weyl's theory [36], F(R) gravity [37], mimetic gravity [38], mimetic F(R) gravity [39], Lyra geometry [40] etc. These customized theories of gravitation may be used to study the accelerating expansion of the universe. Amongst these customized theories of gravitation, here, the Lyra geometry is discussed. Soleng [41] shows that the gauge vector $\phi_{i}$ in Lyra's geometry will play either the role of creation field Hoyle [42] (equal to Hoyle's creation field [42]) or cosmological constant. Sen [43], Sen and Dunn [44], Rosen [45] are some well-known researchers who have studied scalar-tensor theory of gravitation on the basis of Lyra geometry. Halford [46, 47] recommended that the cosmological constant in the general theory of relativity and the constant displacement vector field $\phi_{i}$ in Lyra's geometry perform the same role and the scalar-tensor treatment in the framework of Lyra's geometry expect the identical result, within observational limits as predicted by Einstein's theory of relativity. Bhamra [48], Karade and Borikar [49], Rahaman [50], Khadekar and Nagpure [51], Rahaman et al. [52], Casana et al. [53, 54], Mohanty et al. [55, 56], Shchigolev [57], Mahanta et al. [58], Darabi et al. [59], Mollah et al. [60], Mollah and Priyokumar [61] etc. are the some authors who studied Einstein's field equations of gravitation in the framework of Lyra's geometry and obtained their solutions effectively under different circumstances.\\
\\
The equation of state in relativity and cosmology, which is nothing but the relationship among combined matter, temperature, pressure, energy and energy density for any region of space, plays an important role. Many researchers like Ivanov [62], Sharma and Maharaj [63], Thirukkanesh and Maharaj [64], Feroze and Siddiqui [65], Varela et al. [66] etc. studied cosmological models with linear and non-linear equation of state. For the study of dark energy and general relativistic dynamics in different cosmological models, the quadratic equation of state plays an important role. Dark energy universe with different equations of state were already discussed by various authors like, Bamba et al [25], Nojiri and Odintsov [67, 68], Nojiri et al. [69], Nojiri and Odintsov [70] and Capozziello et al. [71]. The general form of quadratic equation of state

\[p = p_{0} + \alpha \rho + \beta \rho^{2} \ , \]
where $p_{0} , \alpha , \beta$ are the parameters, is nothing but the first term of the Taylor's expansion of an equation of state of the form $p = p(\rho)$ about $\rho = 0$.\\
\\
Considering a non-linear equation of state (EoS) in the form of a quadratic equation $p = p_{0} + \alpha \rho + \beta \rho^{2}$, Ananda $ \&$ Bruni [72] investigated the general relativistic dynamics of Robertson-Walker models. They have shown that in general relativistic theory setting, the anisotropic behaviour at the singularity obtained in the Brane Scenario can be reproduced. In the general theory of relativity, they have also discussed the anisotropic homogeneous and inhomogeneous cosmological models with the consideration of quadratic equation of state of the form

\[p = \alpha \rho + \frac{\rho^{2}}{\rho_{c}} \ ,\]
and attempted to isotropize the model universe in the initial stages when the initial singularity is approached. In this paper, we have taken the quadratic equation of state of the form
\[p = \alpha \rho^{2} - \rho \ ,\]
is considered, where $\alpha \neq 0$ is a constant quantity, but we can take $p_{0} = 0$ to make our calculations simpler. This will not affect the quadratic nature of the equation of state.\\
\\
Considering an equation of state in quadratic form as  $\frac{p}{c^{2}} = -\frac{4\rho^{2}}{3\rho_{p}} + \frac{\rho}{3} - \frac{4\rho_{\Lambda}}{3}$ , Chavanis [73] investigated a four-dimensional Friedmann-Lemaitre-Roberston-Walker (FLRW) cosmological model unifying radiation, vacuum energy and dark energy. Again, by using a quadratic form of equation of state, Chavanis [74] formulated a cosmological model that describes the early inflation, the intermediate decelerating expansion, and the late-time accelerating expansion of the universe.\\
\\
Many authors like Maharaj et al. [75]; Rahaman, F., et al. [76], Feroze and Siddiqui [65] have studied cosmological models on the basis of equation of state in quadratic form under different circumstances. Recently, V. U. M. Rao et al. [77], Reddy et al. [78], Adhav et al. [79] studied Kaluza-Klein Space-time cosmological models with a quadratic equation of state in general and modified theories of relativity.\\
\\
Inspired by the above studies, here, we have investigated a homogeneous and anisotropic Bianchi type III cosmological model universe with a quadratic equation of state in Lyra Geometry setting and find out the realistic solutions.\\
\\
This paper has been put into order as follows: In Sec. 2, we have formulated the problem where the physical and kinematical parameters are defined. In this section, exact solutions of the field equations are also obtained and the graphs of some of the parameters are shown as well. Sec. 3 consists of the physical and geometrical aspects of the derived model. Concluding remarks are given in Sec. 4. In Sec. 5, an acknowledgement to the funding authority finds a place.

\section{Formulation of problem}

Let us consider the homogeneous and anisotropic space-time metric described by the Bianchi type III line element in the form
\begin{equation}
ds^{2}=A^{2}dx^{2}+B^{2}e^{-2{\alpha}x}dy^{2}+C^{2}dz^{2}-dt^{2}
\end{equation}
where $A$, $B$ and $C$ are the scale factors, which are functions of cosmic time $t$ only and $\alpha \neq 0$ is a constant.\\
\\
The Einstein field equations in Lyra Geometry (as taken by Sen [43]; Sen and Dunn [44]) in geometric units ($c = G = 1$) is given by
\begin{equation}
R_{ij}-\frac{1}{2}Rg_{ij}+\frac{3}{2}\phi_{i}\phi_{j}-\frac{3}{4}g_{ij}\phi^k\phi_k=-8\pi T_{ij}
\end{equation}
and the energy momentum tensor $T_{ij}$ is taken as

\begin{equation}
T^{i}_{j}=(\rho + p) u^{i}u_{j}+pg^{i}_{j}
\end{equation}
where, $u^{i} = (0, 0, 0, 1)$ is the four velocity vector, $\phi_{i}=(0, 0, 0, \beta(t))$ is the displacement vector; $R_{ij}$ is the Ricci tensor; $R$ is the Ricci scalar, $\rho$  is the energy density and $p$ is the pressure so that

\begin{equation}
g_{ij}u^{i}u^{j}=1
\end{equation}
\\
Therefore, we have
\begin{equation}
T^{1}_{1}=T^{2}_{2}=T^{3}_{3} = p ~~ ; ~~ T^{4}_{4} = - \rho ~~~~ \textrm{and} ~~~~ T^{i}_{j}=0  ~~~ \textrm{for all} ~~ i\neq 0
\end{equation}
\\
Also, the general quadratic form of the Equation of state (EoS) for the matter of distribution is given by

\[p=p(\rho)= a\rho^{2} + b\rho + c\]
where $a \neq 0$, $b$ and $c$ are constants.\\
\\
From the field equations (2), the continuity equation is given by

\begin{equation}
\dot{\rho} + \frac{3}{2}\beta \dot{\beta} + \left[(\rho + p) + \frac{3}{2}\beta^{2}\right] \left(\frac{\dot{A}}{A}+\frac{\dot{B}}{B}+\frac{\dot{C}}{C}\right)= 0
\end{equation}
\\
Again, by the use of the energy conservation equation  $T^{i}_{j;i}$ , the continuity equation for matter can be written as

\begin{equation}
\left(R^{i}_{j}-\frac{1}{2}g^{i}_{j}R\right)_{;i} + \frac{3}{2}\left( \phi^{i}\phi_{j} \right)_{;i}-\frac{3}{4}\left( g^{i}_{j}\phi^{k}\phi_{k} \right)_{;i} = 0
\end{equation}
\\
Simplifying equation (7) we have

\begin{equation}
\frac{3}{2}\phi_{j}\left[ \frac{\partial\phi^{i}}{\partial x^{i}} + \phi^{l}\Gamma^{i}_{li} \right] + \frac{3}{2}\phi^{i}\left[ \frac{\partial\phi_{j}}{\partial x^{i}} - \phi_{l}\Gamma^{l}_{ij} \right] - \frac{3}{4}g^{i}_{j}\phi_{k}\left[ \frac{\partial\phi^{k}}{\partial x^{i}} + \phi^{l}\Gamma^{k}_{lj} \right] - \frac{3}{4}g^{i}_{j}\phi^{k}\left[ \frac{\partial\phi_{k}}{\partial x^{i}} - \phi_{l}\Gamma^{l}_{kj} \right] = 0
\end{equation}
\\
This equation (8) identically satisfied for $j = 1, 2, 3$.\\
\\
But, for $j = 4$, this equation reduces to

\begin{equation}
\frac{3}{2}\beta \dot{\beta} + \frac{3}{2}\beta^{2} \left(\frac{\dot{A}}{A}+\frac{\dot{B}}{B}+\frac{\dot{C}}{C}\right)= 0
\end{equation}
\\
Therefore, the continuity equation (6) reduces to

\begin{equation}
\dot{\rho} + (\rho + p) \left(\frac{\dot{A}}{A}+\frac{\dot{B}}{B}+\frac{\dot{C}}{C}\right)= 0
\end{equation}
\\
Therefore, in a comoving coordinate system, the Bianchi type III space-time metric (1) for the energy momentum tensor (3), the Einstein's field equations (2) reduces to

\begin{equation}
\frac{\ddot{B}}{B}+\frac{\ddot{C}}{C}+\frac{\dot{B}\dot{C}}{BC}+\frac{3}{4}\beta^{2}=-8 \pi p
\end{equation}

\begin{equation}
\frac{\ddot{A}}{A}+\frac{\ddot{C}}{C}+\frac{\dot{A}\dot{C}}{AC}+\frac{3}{4}\beta^{2}=-8 \pi p
\end{equation}

\begin{equation}
\frac{\ddot{A}}{A}+\frac{\ddot{B}}{B}+\frac{\dot{A}\dot{B}}{AB}-\frac{\alpha^{2}}{A^{2}}+\frac{3}{4}\beta^{2}=-8 \pi p
\end{equation}

\begin{equation}
\frac{\dot{A}\dot{B}}{AB}+\frac{\dot{A}\dot{C}}{AC}+\frac{\dot{B}\dot{C}}{BC}-\frac{\alpha^{2}}{A^{2}}-\frac{3}{4}\beta^{2}=-8 \pi \rho
\end{equation}

\begin{equation}
\alpha \left( \frac{\dot{A}}{A} - \frac{\dot{B}}{B} \right) = 0
\end{equation}
where the overhead dot denote derivatives with respect to the cosmic time $t$.\\
\\
Now, for the metric (1) , important physical quantities like Volume $V$, average Scale factor $R$ , Expansion Scalar $\theta$ , Hubble's parameter $H$ , Shear Scalar $\sigma$ , Anisotropy Parameter $\Delta$ and Deceleration parameter $q$ are defined as

\begin{equation}
V = R^{3} = ABCe^{-\alpha x}
\end{equation}

\begin{equation}
\theta = \frac{\dot{A}}{A}+\frac{\dot{B}}{B}+\frac{\dot{C}}{C}
\end{equation}

\begin{equation}
H = \frac{1}{3}\theta = \frac{1}{3}\left(\frac{\dot{A}}{A}+\frac{\dot{B}}{B}+\frac{\dot{C}}{C}\right)
\end{equation}

\begin{equation}
\sigma^{2} = \frac{1}{3}\left( \frac{\dot{A}^{2}}{A^{2}} + \frac{\dot{B}^{2}}{B^{2}} + \frac{\dot{C}^{2}}{C^{2}}
            - \frac{\dot{A}\dot{B}}{AB} - \frac{\dot{A}\dot{C}}{AC} - \frac{\dot{B}\dot{C}}{BC}\right)
\end{equation}

\begin{equation}
\Delta = \frac{1}{3H^{2}}\left( \frac{\dot{A}^{2}}{A^{2}} + \frac{\dot{B}^{2}}{B^{2}} + \frac{\dot{C}^{2}}{C^{2}} \right) -1
\end{equation}
and

\begin{equation}
q = 3 \frac{d}{dt}\left(\frac{1}{\theta}\right) - 1
\end{equation}
\\
Since $\alpha \neq 0$ , so from (15), we have

\[ \frac{\dot{A}}{A} = \frac{\dot{B}}{B} \]
\\
Integrating it, we get

\begin{equation}
A = kB
\end{equation}
where $k$ is a constant. Without loss of generality, we may choose $k=0$ then we have
\begin{equation}
A = B
\end{equation}
\\
Therefore, the equations (11)-(14) reduces to
\begin{equation}
\frac{\ddot{B}}{B}+\frac{\ddot{C}}{C}+\frac{\dot{B}\dot{C}}{BC}+\frac{3}{4}\beta^{2}=-8 \pi p
\end{equation}

\begin{equation}
2\frac{\ddot{B}}{B}+\frac{\dot{B}^{2}}{B^{2}}-\frac{\alpha^{2}}{B^{2}}+\frac{3}{4}\beta^{2}=-8 \pi p
\end{equation}

\begin{equation}
\frac{\dot{B}^{2}}{B^{2}}+2\frac{\dot{B}\dot{C}}{BC}-\frac{\alpha^{2}}{B^{2}}-\frac{3}{4}\beta^{2}= 8 \pi \rho
\end{equation}
\\
The equations (24) - (26) represents a system of three independent simultaneous equations involving five unknown parameters viz. $B$, $C$, $\beta$, $\rho$ and $p$. So, in order to find exact solution of the above system, it is required two more physical conditions involving these parameters. These two conditions are taken as follows-\\

	i)	choosing $b = -1$ and $c = 0$ in general form of quadratic equation of state, we have considered the equation of state in the form

        \begin{equation}
        p = a\rho^{2} - \rho
        \end{equation}
	   where $a$ is a constant and

    ii) we assume that the expansion scalar $\theta$  is proportional to the shear tensor $\sigma^{1}_{1}$ so that we get

        \begin{equation}
        BC = A^{n}
        \end{equation}
        where $n$ is a constant.\\
\\
So, from equations (23) - (26) and (28), it may be obtained

\begin{equation}
n\frac{\ddot{B}}{B} + (n-1)^{2}\frac{\dot{B}^{2}}{B^{2}} + \frac{3}{4}\beta^{2} = -8 \pi p
\end{equation}

\begin{equation}
2\frac{\ddot{B}}{B} + \frac{\dot{B}^{2}}{B^{2}} - \frac{\alpha^{2}}{B^{2}} + \frac{3}{4}\beta^{2} = -8 \pi p
\end{equation}

\begin{equation}
(2n-1)\frac{\dot{B}^{2}}{B^{2}} - \frac{\alpha^{2}}{B^{2}} - \frac{3}{4}\beta^{2} = 8 \pi \rho
\end{equation}
\\
Subtracting (30) from (29) and adding (30) $\&$ (31), we have

\begin{equation}
(n-2)\frac{\ddot{B}}{B} + (n^{2}-2n)\frac{\dot{B}^{2}}{B^{2}} + \frac{\alpha^{2}}{B^{2}} = 0
\end{equation}
and

\begin{equation}
\frac{\ddot{B}}{B} + n\frac{\dot{B}^{2}}{B^{2}} - \frac{\alpha^{2}}{B^{2}} = 4 \pi (\rho - p)
\end{equation}
From equations (32) and (33), it can be obtained that

\begin{equation}
\frac{\ddot{B}}{B} + n\frac{\dot{B}^{2}}{B^{2}} = \frac{4 \pi}{n-1} (\rho - p)
\end{equation}
\\
The equations (16) and (34) will give us

\begin{equation}
\dot{V} = \sqrt{m \rho V^{2} + k_{1}}
\end{equation}
where $k_{1}$ and $m = \frac{8 \pi (n+1)}{n-1}$ are integrating constants.\\
\\
Integrating equation (35), we have
\begin{equation}
\int \frac{dV}{\sqrt{m \rho V^{2}} + k_{1}} = t + k_{2}
\end{equation}
where $k_{2}$ is an integrating constant that represent a shift of cosmic time $t$. Therefore, it can be chosen as zero.\\
\\
Equations (10) and (27) will give us

\begin{equation}
\rho = (a logV)^{-1}
\end{equation}
\\
Now, if we take $k_{1} = k_{1} = 0$, then from equations (36), the volume $V$ is obtained as

\begin{equation}
V = exp \left[\left(\frac{9m}{4a}t^{2}\right)^{\frac{1}{3}}\right]
\end{equation}
\\
So, the scale factors $A$, $B$ and $C$ are obtained from equations (16), (23), (28) and (38) as

\begin{equation}
A = B = exp \left[\frac{1}{n+1}\left(\frac{9m}{4a}t^{2}\right)^{\frac{1}{3}}\right] e ^{\frac{\alpha x}{n+1}}
\end{equation}
and
\begin{equation}
C = exp \left[\frac{n-1}{n+1}\left(\frac{9m}{4a}t^{2}\right)^{\frac{1}{3}}\right] e ^{\frac{(n-1)\alpha x}{n+1}}
\end{equation}
\\
Using equations (39) and (40) in equation (1), the geometry of the model universe is given by

\begin{equation}
\begin{split}
ds^{2}=&exp \left[\frac{2}{n+1}\left(\frac{9m}{4a}t^{2}\right)^{\frac{1}{3}}\right] e ^{\frac{2 \alpha x}{n+1}}\left[dx^{2}+e^{-2{\alpha}x}dy^{2}\right]\\
        &+exp \left[\frac{2(n-1)}{n+1}\left(\frac{9m}{4a}t^{2}\right)^{\frac{1}{3}}\right] e ^{\frac{2(n-1)\alpha x}{n+1}}dz^{2}-dt^{2}
\end{split}
\end{equation}
\\
The use of equation (38) in Equation (37), the energy density $\rho$ is obtained as

\begin{equation}
\rho = \frac{1}{a} \left(\frac{9m}{4a}\right)^{-\frac{1}{3}}t^{-\frac{2}{3}}
\end{equation}
\\
Therefore, from equation (27), the pressure $p$ can be obtained as

\begin{equation}
p = \frac{1}{a} \left(\frac{9m}{4a}\right)^{-\frac{1}{3}}t^{-\frac{2}{3}}\left[ \left(\frac{9m}{4a}\right)^{-\frac{1}{3}}t^{-\frac{2}{3}} -1 \right]
\end{equation}
\\
From equation (31), the displacement vector $\beta$ is given by

\begin{equation}
\frac{3}{4}\beta^{2} = \frac{n-2}{n-1}\frac{4\pi}{a} \left(\frac{9m}{4a}\right)^{-\frac{1}{3}}t^{-\frac{2}{3}}\left[ \frac{2}{n+1} - \left(\frac{9m}{4a}\right)^{-\frac{1}{3}}t^{-\frac{2}{3}} \right]
\end{equation}
\\
For the model universe (41), the other physical and geometrical properties like expansion scalar $\theta$, Hubble's expansion factor $H$, shear scalar $\sigma$, anisotropy parameter $\Delta$ and deceleration parameter $q$ can be easily obtained from equations (17)-(21).
\begin{equation}
\theta = \frac{2}{3} \left(\frac{9m}{4a}\right)^{\frac{1}{3}}t^{-\frac{2}{3}}
\end{equation}
\begin{equation}
H = \frac{2}{9} \left(\frac{9m}{4a}\right)^{\frac{1}{3}}t^{-\frac{2}{3}}
\end{equation}
\begin{equation}
\sigma^{2} = \frac{4(n-2)^{2}}{27(n+1)^{2}} \left(\frac{9m}{4a}\right)^{\frac{2}{3}}t^{-\frac{2}{3}}
\end{equation}
\begin{equation}
\Delta = \frac{2(n-2)^{2}}{(n+1)^{2}} = \textrm{constant} \neq 0 ~~ \textrm{for} ~~ n \neq 2 ~~ \textrm{and} ~~ n \neq -1
\end{equation}
\begin{equation}
q = \frac{3}{2} \left(\frac{9m}{4a}\right)^{-\frac{1}{3}}t^{-\frac{2}{3}} -1
\end{equation}
It is well known that the different values of the parameters will give rise different graph, so the variations of some parameters are shown, by taking particular values of the integrating constants  as  $a = 10$ and $n = 10$ so that $m = 30.730159$, in the following Fig. 1-6.
\begin{figure}[H]
\centerline{\psfig{file=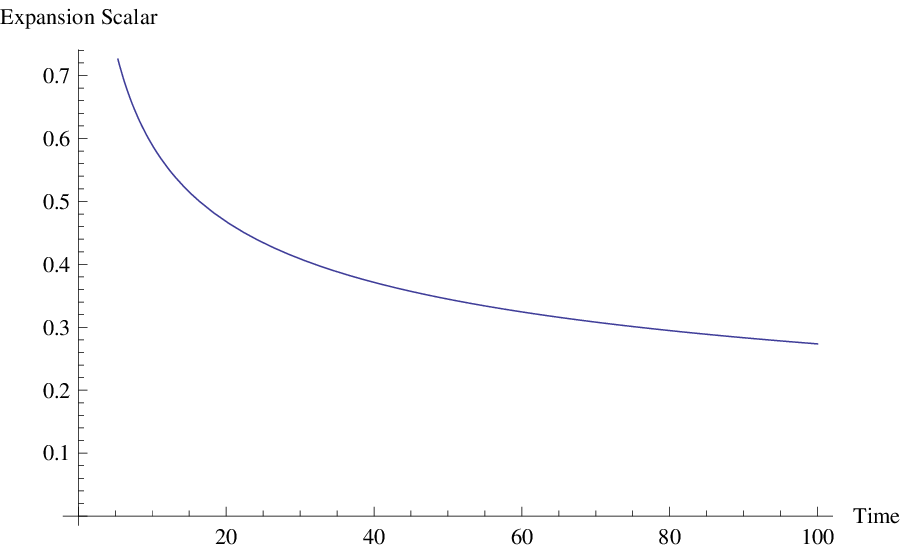,width=3in}}
\vspace*{8pt}
\caption{The variation of Expansion Scalar $\theta$ vs Time $t$, when $a$ = 10 and $n$ = 10 so
that $m$ = 30.730159.\label{Fig. 1.}}

\centerline{\psfig{file=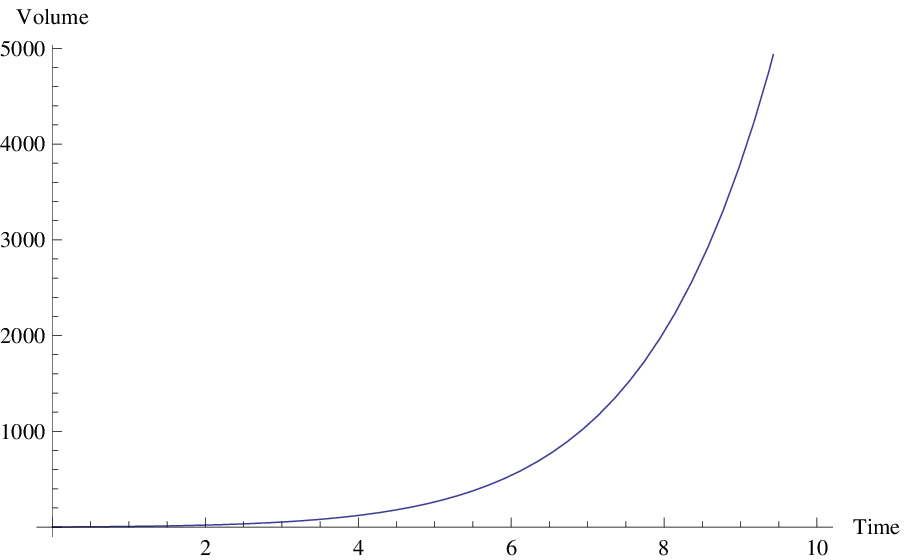,width=3in}}
\vspace*{8pt}
\caption{The variation of Volume $V$ vs Time $t$, when $a$ = 10 and $n$ = 10 so that
$m$ = 30.730159.\label{Fig. 2.}}

\centerline{\psfig{file=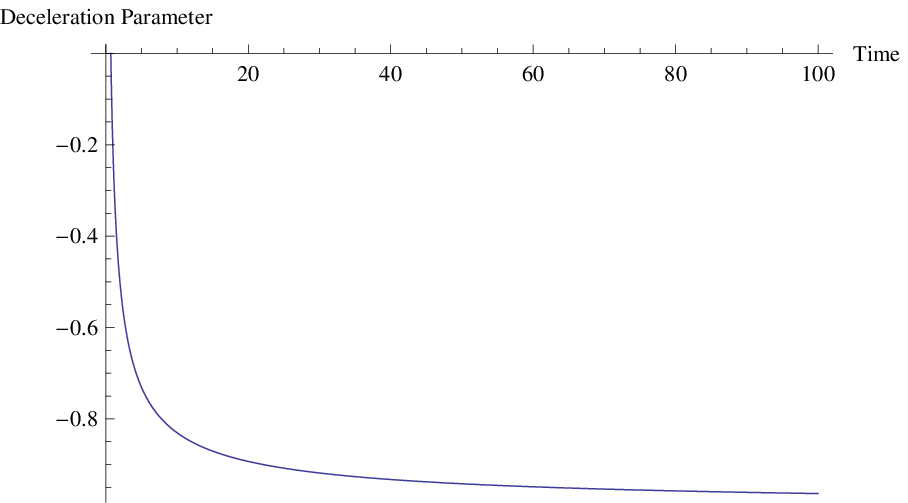,width=3in}}
\vspace*{8pt}
\caption{The variation of Deceleration Parameter $q$ vs Time $t$, when $a$ = 10 and
$n$ = 10 so that $m$ = 30.730159.\label{Fig. 3.}}
\end{figure}

\begin{figure}[H]
\centerline{\psfig{file=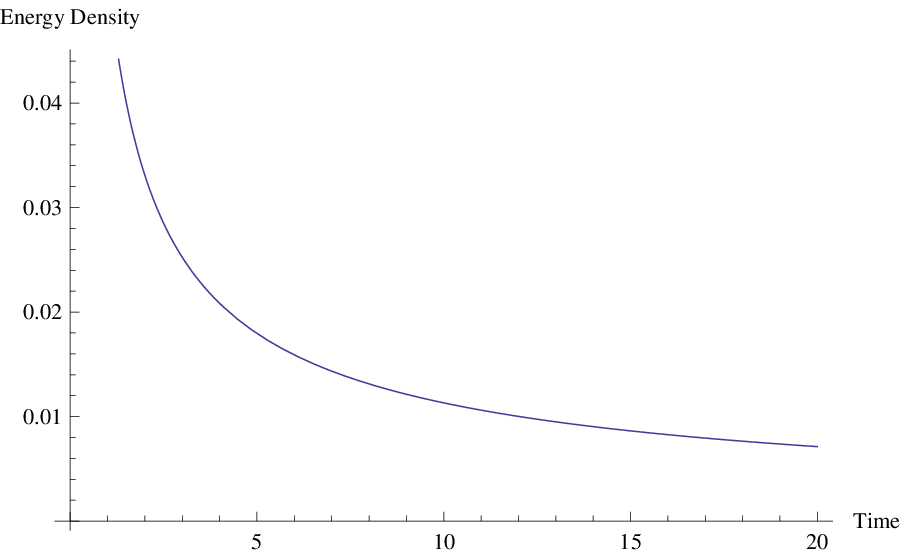,width=3in}}
\vspace*{8pt}
\caption{The variation of Energy Density $\rho$ vs Time $t$, when when $a$ = 10 and $n$ = 10
so that $m$ = 30.730159.\label{Fig. 4.}}

\centerline{\psfig{file=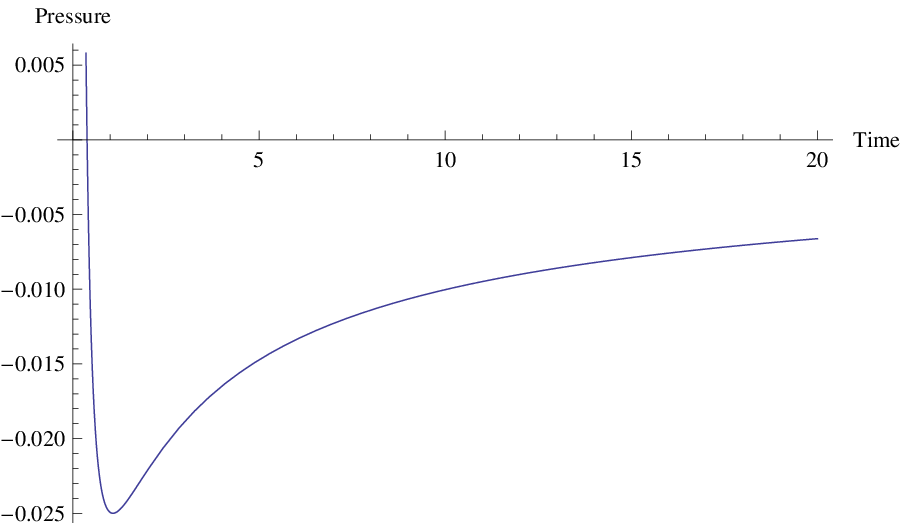,width=3in}}
\vspace*{8pt}
\caption{The variation of Pressure $p$ vs Time $t$, when when $a$ = 10 and $n$ = 10 so
that $m$ = 30.730159.\label{Fig. 5.}}

\centerline{\psfig{file=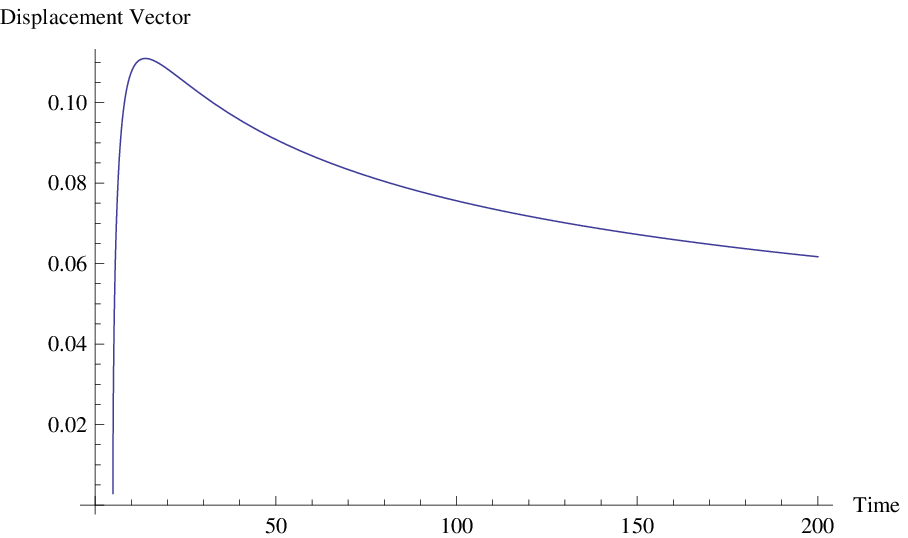,width=3in}}
\vspace*{8pt}
\caption{The variation of Displacement Vector $\beta$ vs Time $t$, when $a$ = 10 and $n$ = 10
so that $m$ = 30.730159.\label{Fig. 6.}}
\end{figure}
\section{Physical and Geometrical Properties of the Model Universe}
The evolution of expansion scalar $\theta$ has been shown in Fig. 1 corresponding to the equation (45) and it is observed that the expansion scalar $\theta$ starts with infinite value at initial epoch of cosmic time $t = 0$. But, as time $t$ progresses, it decreases and becomes constant after some finite time that explains the Big-Bang scenario. From equation (38) and Fig. 2, it is seen that the value of volume $V$ of the model universe is an increasing function of cosmic time $t$ and increases from zero at initial epoch of time to infinite volume whenever $t \rightarrow \infty$ , so our model represents an expanding universe.\\
\\
Now, from equation (49) and Fig. 3, it is clear that initially the value of deceleration parameter $q$ is negative and tends to $-1$ at infinite time i.e. the values of $q$ is in the range $-1 \leq  q \leq 0$ implying that the values of $q$ satisfies the present observational data like Riess et al. [6] and Perlmutter et al. [7]. Also, from the expression (46) for Hubble's expansion factor $H$ it has been observed that $dH/dt$  is negative, so our model universe expands with an accelerated rate.\\
\\
Since the present day universe is isotropic, it is important to observe whether our models evolve into an isotropic or an anisotropic one. In order to investigate the isotropy of the model universe, here, we have considered the simple anisotropy parameter $\Delta$. From equation (48), it has been observed that the anisotropy parameter $\Delta$ is independent of cosmic time $t$ and $\Delta = 0$ for $n = 2$ but whenever $n \neq 2$ and $n \neq 1$ then $\Delta \neq 0$ , which shows that our model universe is isotropic throughout the evolution whenever $n = 2$ but for $n \neq 2$ and $n \neq 1$, the model remains anisotropic throughout the evolution.\\
\\
Again, it is known that the energy conditions for Bianchi type III model is energy density $\rho$ is positive i.e. $\rho > 0$. From Fig. 4 of equation (42), it is seen that the energy $\rho$ is always positive. Also, the Fig. 5 showing the evolution of pressure $p$ depicts that initially when $t \rightarrow \infty$ then $p$ is positive but as the time progresses $p$ changes sign from positive to negative. Therefore, our model universe trespasses through the transition from matter dominated period to inflationary period.\\
\\
From equation (44), the displacement vector $\beta$ is found to be positive which increases rapidly at initial epoch of time but with the increase of time it decreases and at infinite time the displacement vector $\beta$ becomes a small positive constant. The variation of the parameter 'displacement vector $\beta$' is shown in Fig. 6. Also, from equation (47) it can be seen that the shear scalar $\sigma$ is a decreasing function of cosmic time $t$ and vanishes as $t \rightarrow \infty$ . So, our model represents a shear free dark energy cosmological model universe for large values of cosmic time $t$.
\section{Conclusion}	
Investigating a homogeneous and anisotropic space-time described by Bianchi type III metric in presence of perfect fluid in Lyra geometry setting under the assumption of quadratic equation of state (EoS), exact solutions of the Einstein's field equations have been obtained. Here, we have got a model universe which is expanding with acceleration that also passes through the transition from matter dominated period to inflationary period. The displacement vector $\beta$ and shear scalar $\rho$ becomes zero as $t \rightarrow \infty$. So, our model represents a shear free dark energy cosmological model universe for large values of cosmic time $t$. \\
\section{Acknowledgements}

The authors are very thankful to the UGC, India for funding under the sanction Order No. F. 5-332/2014-15/MRP/NERO/2386 to carry out this work successfully.
\\

\end{document}